\documentclass[prc,floatfix,groupedaddress,nofootinbib,showpacs,preprintnumbers,
amsmath,amssymb,amsfonts,superscriptaddress,widetable] {revtex4-1}
\usepackage{graphicx}
\usepackage{dcolumn}
\usepackage{mathrsfs}
\usepackage{bm}
\usepackage{dsfont}
\usepackage{graphicx}
\usepackage{dcolumn}
\usepackage[usenames]{color}
\begin{document}

\title{Relativistic mean field plus exact pairing approach to open shell nuclei}
\author{Wei-Chia Chen}
\email{wc09c@my.fsu.edu} 
\affiliation{Department of Physics, Florida State University, Tallahassee, FL 32306} 
\author{J. Piekarewicz}
\email{jpiekarewicz@fsu.edu}
\affiliation{Department of Physics, Florida State University, Tallahassee, FL 32306}
\author{A. Volya}
\email{volya@physics.fsu.edu}
\affiliation{Department of Physics, Florida State University, Tallahassee, FL 32306} 
\date{\today}
\begin{abstract}
\begin{description}
\item[Background] Pairing correlations play a critical role in determining numerous properties of open-shell nuclei. 
Traditionally, they are included in a mean-field description of atomic nuclei through the approximate 
Bardeen-Cooper-Schrieffer or Hartree-Fock-Bogoliubov formalism. 
\item[Purpose] We propose a new hybrid ``\emph{relativistic-mean-field--plus--pairing}'' approach in which pairing is 
treated exactly so the number of particles is conserved. To verify the reliability of the formalism, we apply it to the 
study of both ground-state properties and isoscalar monopole excitations of the Tin isotopes.
\item[Methods] Accurately-calibrated relativistic mean-field models supplemented by an exact treatment of pairing correlations
are used to compute ground-state observables along the isotopic chain in Tin. In turn, ground-state densities are used
as input to the calculation of giant monopole resonances through a constrained-relativistic approach. 
\item[Results] We compute a variety of ground-state observables sensitive to pairing correlations as well as
the evolution of giant monopole energies along the isotopic chain in Tin. Whereas ground-state properties
are consistent with experiment, we find that pairing correlations have a minor effect on the giant monopole energies.
\item[Conclusions] A new mean-field--plus--pairing approach is introduced to compute properties of open-shell nuclei.
The formalism provides an efficient and powerful alternative to the computation of both ground-state properties and 
monopole energies of open-shell nuclei. We find ground-state properties to be well reproduced in this approach. However, 
as many have concluded before us, we find that pairing correlations are unlikely to provide an answer to the question 
of  ``\emph{why is Tin so soft?}''
\end{description}
\end{abstract}
\maketitle

\section{Introduction}
\label{intro}

Since the early days of nuclear physics, there has been ample experimental evidence in support 
of nuclear pairing in atomic nuclei\,\cite{Mayer:1950,Dean:2003}. Indeed, already in 1950 Maria 
Goeppert Mayer suggested that an even (odd) number of identical nucleons occupying the same 
single-particle orbit of angular momentum $j$ will couple to a total angular momentum of $J\!=\!0$ 
($J\!=\!j)$\,\cite{Mayer:1950}. One remarkable consequence of such an assumption is that all 
even-even nuclei are predicted (and so far observed) to have a ground state with a total angular 
momentum of $J\!=\!0$ and a ground-state energy that is significantly lower relative to that of 
its odd-nucleon neighbors. 

Pairing correlations involve the binding of identical nucleons moving in time-reversed orbits around 
the Fermi surface. In general, the imprint of pairing correlations is observed in a variety of nuclear 
properties, such as binding energies, one nucleon separation energies, single-particle occupancies, 
excitation spectra, level densities, moments of inertia, and low-lying collective modes, among 
others\,\cite{BohrII:1998}. In recent years, the focus of nuclear structure has shifted from the valley 
of stability to the boundaries of the nuclear landscape. Indeed, it is now possible to both synthesize 
and probe the structure of \emph{exotic nuclei}, particularly  neutron-rich and neutron-deficient 
nuclei\,\cite{Dobaczewski:1996,Pfutzner:2012}. Moreover, the ongoing quest for 
\emph{superheavy elements} continues. This quest involves a sustained effort on both their 
synthesis and on theoretical predictions of novel shell structure and new magic 
numbers\,\cite{Hofmann:2000,Oganessian:2010}. It is widely recognized that pairing correlations 
play a critical role along these new frontiers.

Theoretical treatments on nuclear pairing can be classified into two groups: one approximate and the 
other exact. The approximate approaches followed the seminal work of Bardeen, Cooper, and Schrieffer
(BCS)\,\cite{Bardeen:1957} on superconductivity in condensed-matter physics that were extended 
shortly after to the nuclear domain by\,\citet{Bohr:1958}, \citet{Belyaev:1959}, and\,\citet{Migdal:1959}. 
Since then, methods combining a Hartree-Fock formalism with BCS theory (HF+BCS) have been 
developed and widely implemented\,\cite{Tondeur:1979,Nayak:1995}. Although quite successful for 
macroscopic systems, BCS theory suffers from two main disadvantages when applied to finite nuclei. 
First, the BCS formalism does not conserve number of particles. This is not a serious issue for macroscopic 
systems containing $10^{23}$ particles, but it certainly becomes relevant for small systems like atomic 
nuclei. Second, for nuclei whose single-particle-energy spacing around the Fermi surface is greater 
than the typical pairing strength, the BCS formalism generates trivial solutions. These drawbacks
complicate the identification of weakly bound nuclei near the drip line, as the whole concept of drip 
line becomes unclear when the exact number of particles is unknown.
A more sophisticated approach that incorporates pairing correlations in a mean-field framework is the 
Hartree-Fock-Bogoliubov (HFB) formalism\,\cite{Bogoliubov:1959}. In the HFB approach the (short-range) 
particle-particle channel associated with pairing correlations is treated on equal footing as the (long-range) 
particle-hole channel associated with the conventional HF description\,\cite{Ring:2004}. This technique has 
been successfully applied to stable and weakly bound nuclei in both the 
non-relativistic\,\cite{Dobaczewski:1996,Duguet:2001a,Duguet:2001b} and relativistic 
domains\,\cite{Paar:2002gz,Vretenar:2005zz}. However, the violation of particle number remains an 
important drawback of the HFB formalism. To overcome such difficulty, a complicated prescription
is often invoked to project out the state containing the desired number of 
particles\,\cite{Lipkin:1960,Nogami:1964}. Unfortunately, such a prescription along with many other 
ideas have been met with limited success. Meanwhile, a variety of new approaches aimed to solve 
the pairing problem \emph{exactly} were proposed, primarily 
by\,\citet{Richardson:1963a,Richardson:1963b,Richardson:1965}; see also 
Ref.\,\cite{Dukelsky:2004re} and references contained therein. For example, in the Richardson method 
the large-scale diagonalization of a many-body Hamiltonian in a truncated Hilbert space is reduced to a 
set of coupled algebraic equations with a dimension equal to the number of valence particles. Moreover, 
exact solutions to a generalized pairing problem using sophisticated mathematical tools, such as an 
infinite-dimensional algebra\,\cite{Pan:1997rw}, have also been obtained. However, due to their intrinsic complexity 
these formal methods, although exact, are difficult to implement in realistic situations. Perhaps the most promising 
method to solve the pairing problem exactly is the one based on \emph{quasispin symmetry}, first discovered by 
Racah in the 1940s\,\cite{Racah:1942,Racah:1943}. By exploiting the underlying quasispin symmetry, it is possible to 
express the general pairing Hamiltonian in terms of quasispin operators that are far easier to cope with and 
manipulate\,\cite{Kerman:1961pr,Chen:1995}. The formalism was pushed one step 
further in Ref.\,\cite{Volya:2001} by transforming from the quasispin scheme into the seniority scheme, where 
the physical picture becomes clearer and the simplicity and practicality of the method were explicitly demonstrated. 
Further, it was suggested that a self-consistent approach based on the combination of a mean field plus exact 
pairing formalism represents a promising alternative to large-scale diagonalization\,\cite{Volya:2002}. It is precisely 
the goal of the present contribution to implement and examine the power of this promising alternative.
 
In this work we introduce a new hybrid approach to study the properties of open-shell nuclei. The approach is based 
on the combination of an accurately-calibrated relativistic mean field (RMF) model and an exact treatment 
of pairing correlations. In the RMF theory the underlying nucleon-nucleon (NN) interaction is mediated by various 
``mesons" of different spin, parity, and isospin\,\cite{Serot:1984ey,Mueller:1996pm,Serot:1997xg}. With ever increasing 
sophistication, the RMF theory has been extremely successful in describing ground-state properties of even-even nuclei 
all throughout the nuclear chart\,\cite{Lalazissis:1996rd,Lalazissis:1999}. Pairing correlations have been incorporated into 
the RMF approach by adopting either a  BCS or HFB formalism; see 
Refs.\,\cite{Paar:2002gz,Vretenar:2005zz} and references contained therein. However, these approaches inevitably 
suffer from the aforementioned difficulties related to the violation of particle number. To circumvent this problem we 
propose the exact pairing (EP) approach of Ref.\,\cite{Volya:2001} to address the physics of open-shell nuclei. The 
combination of RMF plus EP (``RMF+EP") is both natural and straightforward to implement. Indeed, single-particle 
energies generated from the RMF approximation are the only input required by the EP algorithm to predict the
occupancies of the valence orbitals. In turn, these new (fractional)  occupancies modify the resulting single-particle 
spectrum---which is then fed back into the EP algorithm. This process continues until self-consistency is achieved.

We illustrate the power and utility of this combined approach by computing ground-state properties and giant-monopole 
energies for the Tin isotopes. With an assumed closed-shell structure for both ${}^{100}$Sn and ${}^{132}$Sn, 
the Tin isotopes serve as a good arena for examining pairing correlations. In particular, we examine a few ground-state 
observables that highlight the critical role of pairing correlations, such as the odd-even staggering in the neutron separation 
energy. However, we are particularly interested in examining the impact (if any!) of pairing correlations on the softening of the 
isoscalar giant monopole resonance (GMR). The GMR, also known as the nuclear \emph{breathing mode}, is a radial density 
oscillation that provides a unique access to the experimentally inaccessible incompressibility of neutron-rich matter---a 
fundamental property of the equation of state. The distribution of isoscalar monopole strength has been traditionally 
measured using inelastic $\alpha$-scattering at very small angles\,\cite{Harakeh:2001}. Indeed, the distribution of 
monopole strength has been measured in ${}^{90}$Zr, ${}^{116}$Sn, ${}^{144}$Sm, and 
${}^{208}$Pb\,\cite{Youngblood:1999,Lui:2004wm,Uchida:2003, Uchida:2004bs} and, with the possible exception of 
${}^{116}$Sn, is accurately reproduced by mean-field plus random-phase-approximation (RPA) calculations. However, 
more recent experimental studies of GMR energies along the isotopic chains in both Tin\,\cite{Li:2007bp,Li:2010kfa} and 
Cadmium\,\cite{Patel:2013} have revealed that the softening observed in ${}^{116}$Sn is endemic to both isotopic
chains\,\cite{Piekarewicz:2009gb,Piekarewicz:2013bea}. A popular explanation behind this anomaly is the critical role 
that pairing correlations play in the physics of these superfluid nuclei\,\cite{Li:2008hx,Khan:2009xq,
Khan:2009ih,Khan:2010mv,Vesely:2012dw}. Although the conclusions have been mixed and seem to depend on the 
character of the pairing force, it appears that pairing correlations are unlikely to provide a definite answer to the question 
of \emph{why is Tin so soft?}  Here too we address this critical question within the context of the RMF+EP approach. 
In particular, we apply the RMF+EP approximation to calculate the relevant ground-state properties required to compute 
the centroid energies of the Tin isotopes from the recently implemented constrained-RMF approach\,\cite{Chen:2013tca}. 
We conclude, as many have done before us, that pairing correlations have a minor effect on the GMR energies 
of the Tin isotopes.

The manuscript has been organized as follows. In Sec.\,\ref{Formalism} we outline separately the RMF and EP approaches. 
Although the description and implementation of both of these techniques have been discussed in detail elsewhere 
\cite[see Refs.][and references contained therein]{Volya:2001,Todd-Rutel:2005fa} a brief review is provided in an
effort to maintain the manuscript self-contained. In particular, Sec.\,\ref{Formalism} puts special emphasis on the implementation 
of the EP approach on top of an RMF approximation. In Sec.\,\ref{Results} we display results obtained with the newly developed
RMF+EP approach for some selective set of ground-state observables and GMR energies along the isotopic chain in Tin. 
Finally, we offer our conclusions in Sec.\,\ref{Conclusions}.

\section{Formalism}
\label{Formalism}
In this section we briefly outline the formalism required to compute ground-state properties and GMR energies
in a RMF+EP approach. We start by reviewing the general features of the RMF theory and then proceed to a discussion
of the exact-pairing approach and how to merge them together. We finish this section by describing how the RMF+EP 
framework may serve as input to the constrained-RMF approach to compute GMR energies.

\subsection{Relativistic Mean Field Theory}
\label{RMFT}

In the RMF theory a nucleus is described in terms of protons and neutrons interacting through the exchange of ``mesons"
of various spins, parities, and isospins. The interactions among the particles can be described by an effective Lagrangian 
density given as follows\,\cite{Walecka:1974qa,Serot:1984ey,Mueller:1996pm,Serot:1997xg,Horowitz:2000xj}:
\begin{eqnarray}
{\mathscr L}_{\rm int} &=&
\bar\psi \left[g_{\rm s}\phi   \!-\! 
         \left(g_{\rm v}V_\mu  \!+\!
    \frac{g_{\rho}}{2}{\mbox{\boldmath $\tau$}}\cdot{\bf b}_{\mu} 
                               \!+\!    
    \frac{e}{2}(1\!+\!\tau_{3})A_{\mu}\right)\gamma^{\mu}
         \right]\psi \nonumber \\
                   &-& 
    \frac{\kappa}{3!} (g_{\rm s}\phi)^3 \!-\!
    \frac{\lambda}{4!}(g_{\rm s}\phi)^4 \!+\!
    \frac{\zeta}{4!}   g_{\rm v}^4(V_{\mu}V^\mu)^2 +
   \Lambda_{\rm v}\Big(g_{\rho}^{2}\,{\bf b}_{\mu}\cdot{\bf b}^{\mu}\Big)
                           \Big(g_{\rm v}^{2}V_{\nu}V^{\nu}\Big)\;,
 \label{LDensity}
\end{eqnarray}
where $\psi$ represents the isodoublet nucleon field, $A_{\mu}$ is the photon field, and $\phi$, $V_{\mu}$, and ${\bf b}_{\mu}$ 
are the isoscalar-scalar $\sigma$-, isoscalar-vector $\omega$-, and isovector-vector $\rho$-meson field, respectively. The 
conventional Yukawa couplings between nucleons and mesons appear in the first line of Eq.(\ref{LDensity}). In the original Walecka 
model\,\cite{Walecka:1974qa} it was sufficient to include the two isoscalar mesons to account for the saturation of symmetric nuclear 
matter at the mean-field level. Later on, the model was extended by Horowitz and Serot\,\cite{Serot:1979dc,Horowitz:1981xw} to 
include the isovector $\rho$-meson and the photon. This formulation was successful in reproducing some ground-state properties 
with an accuracy that rivaled some of the most sophisticated non-relativistic formulations of the time. However, in order to further
improve the standing of the model, it was necessary to include nonlinear self and mixed interactions between the mesons;
these nonlinear meson interactions are given on the second line of Eq.(\ref{LDensity}). For example, the introduction of the 
scalar self-interaction ($\kappa$ and $\lambda$) by\,\citet{Boguta:1977xi} reduces the incompressibility coefficient of symmetric
nuclear matter from the unreasonably large value predicted by the Walecka model\,\cite{Walecka:1974qa,Serot:1984ey} to one
that is consistent with measurements of the distribution of isoscalar monopole strength in medium to heavy nuclei.
Moreover, \citet{Mueller:1996pm} found possible to build models with different values for the quartic $\omega$-meson coupling
($\zeta$) that reproduced the same nuclear properties at normal densities (such as the incompressibility coefficient) but that
produced maximum neutron-star masses that differ by almost one solar mass. Hence, $\zeta$ can be used to efficiently tune 
the maximum neutron star mass\,\cite{Demorest:2010bx,Antoniadis:2013pzd}. Finally, the density dependence of the symmetry 
energy, which has important implications from nuclear structure to astrophysics, is governed by the $\omega$-$\rho$ mixed 
interaction ($\Lambda_{\rm v}$)\,\cite{Horowitz:2000xj,Horowitz:2001ya}. RMF parameters for the two models employed in this
work---FSUGold (or ``FSU" for short)\,\cite{Todd-Rutel:2005fa} and NL3\,\cite{Lalazissis:1996rd}---are given in Table\,\ref{Table1}.
\begin{widetext}
\begin{center}
\begin{table}[h]
\begin{tabular}{|l||c|c|c|c|c|c|c|c|c|c|}
\hline
Model      &  $m_{\rm s}$  &  $m_{\rm v}$  &  $m_{\rho}$  &  $g_{\rm s}^2$  &  $g_{\rm v}^2$  &  $g_{\rho}^2$  
               &  $\kappa$       &  $\lambda$    &  $\zeta$       &   $\Lambda_{\rm v}$  \\
\hline
\hline
FSU         & 491.500 & 782.500 & 763.000 & 112.1996 & 204.5469 & 138.4701 
               & 1.4203  & $+$0.023762 & 0.06 & 0.030  \\
NL3         & 508.194 & 782.501 & 763.000 & 104.3871 & 165.5854 &  79.6000 
               & 3.8599  & $-$0.015905 & 0.00 & 0.000  \\
\hline
\end{tabular}
\caption{Parameter sets for the two accurately calibrated relativistic mean-field models used in the text: 
FSUGold\,\cite{Todd-Rutel:2005fa} and NL3\,\cite{Lalazissis:1996rd}. The parameter $\kappa$ and the 
meson masses $m_{\rm s}$, $m_{\rm v}$, and $m_{\rho}$ are all given in MeV. The nucleon mass has 
been fixed at  $M\!=\!939$~MeV in both models.}
\label{Table1}
\end{table}
\end{center}
\end{widetext}

In the relativistic mean-field limit, the meson fields satisfy (classical) nonlinear Klein-Gordon equations---with the relevant baryon 
densities acting as source terms. In turn, these meson fields provide the (scalar and vector) mean-field potentials for the nucleons. 
Solution of the Dirac equation provide single-particle energies and wave-functions, which are then used to construct the appropriate 
one-body densities that act as sources for the meson fields. This procedure is then repeated until self-consistency is achieved. In
particular, the outcome from such a self-consistent procedure are a variety of ground-state properties, such as the spectrum of 
Dirac orbitals and density profiles. For a detailed description of the formalism and implementation of the RMF approach we refer
the reader to Ref.\,\cite{Todd:2003xs}. We note, however, that the only input required for the implementation of the exact-pairing 
approach are the single-particle energies of the valence orbitals.

\subsection{Exact Solution of the Pairing Problem}
\label{EP}

The RMF theory has been enormously successful in computing ground-state properties and collective excitations of 
even-even nuclei throughout the nuclear chart\,\cite{Lalazissis:1996rd,Lalazissis:1999}. In addition, pairing correlations 
for the description of open-shell nuclei are now routinely incorporated into the relativistic formalism via either a BCS or 
HFB approximation\,\cite[see][and references contained therein]{Vretenar:2005zz}. However, it is the main purpose of 
this work to explore an alternative approach in which the pairing problem is solved exactly\,\cite{Volya:2001}. In 
particular---and in stark contrast to the BCS and HFB approximations---particle number is exactly conserved in this 
formulation.

The general pairing Hamiltonian employed in this manuscript is given by the following expression:
\begin{equation}
H = \sum_{jm} \epsilon_{j} a^{\dagger}_{jm} a_{jm} + 
\frac{1}{4} \sum_{jj'} G_{jj'} \sum_{mm'} a^{\dagger}_{jm} \tilde{a}^{\dagger}_{jm} \tilde{a}_{j'm'} a_{j'm'} \,,
\label{Hpair1}
\end{equation}
where $\epsilon_{j}$ is a set of RMF single-particle energies, $G_{jj'}$ are pairing energies (for $j\!=\!j'$) and pair-transfer 
matrix elements (for $j\!\ne\!j'$). Note that in order to avoid double counting, the monopole contribution to the energy 
will need to be removed, as will be shown below. Nucleon creation and annihilation operators into a single-particle orbit 
labeled by quantum numbers $j$ and $m$ are described by $a^{\dagger}_{jm}$ and $a_{jm}$, respectively. Finally, 
$\tilde{a}_{jm}$ is a time-reversed operator defined as
\begin{equation}
\tilde{a}_{jm} = (-1)^{j-m}a_{j-m} \;.
\end{equation}

The pairing problem with the above Hamiltonian can be solved exactly by introducing quasispin operators for each individual 
single-particle orbital\,\cite{Anderson:1958,Kerman:1961ap}. In particular, the pairing Hamiltonian may be re-written in 
terms of the quasispin operators as 
\begin{equation}
H = \sum_{j}\epsilon_{j}\Omega_{j}+  2\sum_{j} \epsilon_{j}L_{j}^{z}
+ \sum_{jj'} G_{jj'}L_{j}^{+}L_{j'}^{-} \;,
\label{Hpair2}
\end{equation}
where $\Omega_{j}\!=\!(2j\!+\!1)/2$ represents the pair degeneracy of the $j$-orbital and the three
quasispin operators associated to such orbital are defined as follows:
\begin{subequations}
\begin{align}
  L_{j}^{-} =& \frac{1}{2} \sum_{m} \tilde{a}_{jm}a_{jm} = \frac{1}{2}\sqrt{2j+1}
  \sum_{m}\big\langle jm,j,-m\big|00\big\rangle a_{j-m}a_{jm}\,,\\
  L_{j}^{+} =& \frac{1}{2} \sum_{m} a^{\dagger}_{jm} \tilde{a}^{\dagger}_{jm} = \frac{1}{2}\sqrt{2j+1}
  \sum_{m}\big\langle jm,j,-m\big|00\big\rangle a^{\dagger}_{jm}a^{\dagger}_{j-m}\,, \\ 
  L_{j}^{z} =& \frac{1}{2} \sum_{m}\Big(a_{jm}^{\dagger}a_{jm}-\frac{1}{2}\Big)=\frac{1}{2}(N_{j}-\Omega_{j})\,. 
\label{Quasispin}
\end{align}
\end{subequations}
From the above definition it is readily apparent that the operator $L_{j}^{+}$ ($L_{j}^{-}$) creates (destroys) a nucleon 
pair of total angular momentum $J\!=\!0$. Moreover, as the name indicates, the quasispin operators satisfy an SU(2) 
algebra with canonical commutation relations. That is, 
\begin{equation}
 [L_{j}^{+},L_{j'}^{-}]=2\delta_{jj'}L_{j}^{z}
  \quad {\rm and} \quad
 [L_{j}^{z},L_{j'}^{\pm}]=\pm\delta_{jj'}L_{j}^{\pm}\,.
\label{SU2}
\end{equation}
An enormous advantage of introducing the concept of quasispin is that one can bring to bear the full power of the 
angular-momentum algebra into the problem\,\cite{Kerman:1961pr,Chen:1995}. Moreover, one can map the quasispin
basis into the more intuitive \emph{``seniority"} basis that is determined by the seniority quantum number $s_{j}$ and 
the partial occupancy $N_{j}$ of each orbital. Note that the seniority $s_{j}$ of each level $j$ represents the number of 
unpaired particles in such orbital. Given that the pairing Hamiltonian can only transfer pairs of particles, the seniority 
quantum number of each level is conserved. By the same token, the partial occupancy occupancy $N_{j}$ of each orbital
is not conserved. However, in contrast to the BCS and HFB formalism, the total number of particles is exactly conserved 
in this approach. This is one of the major advantages of the EP approach as the exact conservation of the total number of 
particles avoids any reliance on complicated projection prescriptions. Finally, as the mapping between the seniority and 
quasispin bases is straightforward\,\cite{Kerman:1961ap,Racah:1942}, one can evaluate matrix element of the Hamiltonian 
in the seniority basis by first transforming into the quasispin basis and then using the well known properties of the raising 
and lowering  operators\,\cite{Volya:2001}. An overview of the exact-pairing approach and its applications may be found in 
Ref.\,\cite{Zelevinsky:2003}. In addition, a simple illustration of the EP method is given in the appendix.

\subsection{Constrained Relativistic Mean Field Theory}
\label{CRMFT}

One of the central goals of the present manuscript is to investigate the effect of pairing correlations on the GMR energies 
of the Tin isotopes. To do so we implement the newly developed \emph{constrained}-RMF (CRMF) approach introduced in 
Ref.\,\cite{Chen:2013tca}. Although the constrained approach is unable to provide the full distribution of monopole strength, 
it is both accurate and efficient in estimating GMR energies. Indeed, the CRMF formalism that builds on the time-tested 
nonrelativistic formulation has been shown to provide an extremely favorable comparison against the predictions of a 
full relativistic RPA approach\,\cite{Chen:2013tca}. 

The \emph{constrained} GMR energy is defined in terms of two moments of the distribution of strength: 
\begin{equation}
 E_{\rm con} = \sqrt{\frac{m_{1}}{m_{-1}}}, 
 \label{Econ}
\end{equation}
where moments of the isoscalar distribution of monopole ($E0$) strength $R(\omega;E0)$ are given by
\begin{equation}
 m_{n}(E0) \equiv \int_{0}^{\infty} \omega^{n} R(\omega;E0) \ d\omega.
\end{equation}
Note that we distinguish here the constrained energy from the corresponding \emph{centroid} energy that is customarily 
defined as $E_{\rm cen}\!=\!m_{1}/m_{0}$. In particular, assuming a simple Lorentzian distribution of strength one obtains:
\begin{equation}    
 E_{\rm cen}({\rm RPA})= \omega_{0} 
 \quad{\rm and}\quad
 E_{\rm con}({\rm RPA})=
 \sqrt{\omega_{0}^{2}+\Gamma^{2}/4} \;,
  \label{EcenEcon}
\end{equation}
where $\omega_{0}$ is the resonance energy and $\Gamma$ the full width at half maximum. 

The great virtue of the constrained approach is that both of the moments involved in $E_{\rm con}$ may be directly 
computed from ground-state observables. In particular, using Thouless theorem one may compute the $m_1$ moment 
(also known as the energy weighted sum) by evaluating a suitably defined double commutator\,\cite{Harakeh:2001}. 
This procedure yields,
\begin{equation}    
  m_{1}(E0) \equiv \int_{0}^{\infty}\!\omega R(\omega;E0)\,d\omega 
  = \frac{2\hbar^2}{M} A\langle r^2\rangle
  = \frac{2\hbar^2}{M} \int r^{2}\rho({\bf r})\,d^{3}r \;,
 \label{EWSR}
\end{equation}
where $M$ is the nucleon mass and $\rho({\bf r})$ the ground-state baryon density. Similarly, by invoking the 
``\emph{dielectric theorem}" the $m_{-1}$ moment may be written as follows\,\cite{Ring:2004}:
\begin{equation}
 m_{-1}(E0) \equiv \int_{0}^{\infty} \omega^{-1} R(\omega;E0)\,d\omega 
 = -\frac{1}{2} \left[ \frac{d}{d\eta} \int r^2 \rho({\bf r};\eta)\,d^3r \right]_{\eta = 0}.  
 \label{IEWSR}
\end{equation}
Here $\rho({\bf r};\eta)$ is a slightly perturbed ground-state density obtained from solving the RMF equations
by adding a constrained one-body term $V_{\rm con}(r)\!=\!\eta r^2$ to the repulsive vector 
interaction\,\cite{Chen:2013tca}. Such constrained term ``squeezes'' the nucleus making it more compact, 
thereby mimicking the characteristic radial density oscillation of the GMR. Clearly, the constrained approach
is significantly faster than the RPA (or quasiparticle RPA) as the required moments of the distribution are
calculated directly from suitably modified ground-state properties, rather than from the full distribution of
monopole strength. To study the impact of pairing correlations on the GMR energies of the Tin isotopes,
ground-state densities will be computed in the combined RMF+EP formalism, as we describe in the next 
section. 

\subsection{Relativistic Mean Field plus Exact Pairing Formalism}
\label{RMFEP}

Having briefly outlined the RMF theory and the EP method, we now use ${}^{116}$Sn nucleus to illustrate the implementation 
of the combined RMF+EP approach. Since the Tin isotopes have a closed proton shell, we regard the ${}^{100}$Sn nucleus 
as an inert core and then limit the treatment to neutron-neutron ({\it nn}) pairing in the valence shell. In the particular case of 
${}^{116}$Sn, there are 16 valence neutrons residing in a shell consisting of 5 closely spaced orbitals (1g${}_{7/2}$, 2d${}_{5/2}$, 
2d${}_{3/2}$, 3s${}_{1/2}$, and 1h${}_{11/2}$) that can accommodate up to a maximum of 32 neutrons. For an RMF calculation 
without pairing correlations, these 16 neutrons fill up the 1g${}_{7/2}$, 2d${}_{5/2}$, and (half of the) 2d${}_{3/2}$ orbitals; the other 
two orbitals remain completely empty. Such a prescription seems rather unnatural given that the energy difference between filled 
and empty orbitals is comparable to the strength of the pairing interaction. In order to remedy this situation, we invoke pairing
correlations to redistribute the valence particles among the 5 orbitals. To do so, we solve the pairing problem exactly using the 
energy spectrum ({$\epsilon_{j}$}) generated by the RMF model as input to the pairing Hamiltonian.
The Hilbert space for the EP problem is obtained by computing all possible ways to distribute the 8 neutron pairs among these 5 
orbitals. This results in 110 different configurations which serve as the basis for the pairing Hamiltonian. We note that even 
though ${}^{116}$Sn resides in the middle of the shell, the computational demands required to solve the EP problem 
exactly---namely, diagonalizing a $110\!\times\!110$ matrix---are very modest. Given that the pairing strengths in the G-matrix 
approach of Ref.\,\cite{Holt:1998} have been constrained by experiment, the only input required to compute all matrix elements 
of the pairing Hamiltonian are the single-particle energies predicted by the RMF model\,\cite{Volya:2001}. Diagonalizing the pairing 
Hamiltonian mixes all 110 configurations and results in a correlated lowest-energy state that is expressed as a linear combination 
of all these configurations. In particular, this leads to the \emph{fractional occupancy} $\langle N_{j}\rangle$ of each orbital in the 
valence shell; this represents one of the hallmarks of pairing correlations. However, in contrast to other approaches to the pairing 
problem, in the EP formalism the total number of particles is exactly conserved: $N\!\equiv\!\sum_{j}\langle N_{j}\rangle$. Having 
obtained these new fractional occupancies, the RMF problem is solved again, but now with updated baryon densities. These baryon 
densities generate new meson fields, new mean-field potentials, and ultimately a new single-particle spectrum. The updated 
single-particle spectrum $\epsilon_{j}$ now serves as the new input to the EP problem which in turn generates a new set of partial 
occupancies. This iterative procedure is repeated until all fractional occupancies $\langle N_{j}\rangle$  and single-particle energies 
$\epsilon_{j}$ have converged. We note that the RMF+EP approach is self-consistent and particle number is conserved at every 
step in the iterative procedure.

Once the calculation has converged, one must then compute the pairing correlation energy. The correlation energy is obtained
by subtracting from the ground-state energy of the pairing Hamiltonian $E_{0}$ the ``naive'' single-particle contribution. In this way 
the correlation energy accounts for the extra binding energy gained due to pairing. However, given that the diagonal pairing 
strengths $G_{jj}$---corresponding to the monopole part of EP problem---have already been included in the RMF calculation,
one must also remove the monopole energy to avoid double counting. This yields the following form for the correlation 
energy\,\cite{Volya:2001}:
\begin{equation}
 E_{\rm corr} = E_{0} - \sum_{j} \epsilon_{j} \langle N_{j}\rangle - 
 \sum_{j} \frac{G_{jj}}{2\Omega_{j}-1} \frac{\langle N_{j}\rangle \big(\langle N_{j}\rangle-1\big)}{2}\,.
 \label{Ecorr}
\end{equation} 
Ultimately, the correlation energy is added to the corresponding RMF prediction and this is the nuclear binding energy that will 
be reported in Sec.\,\ref{Results}. Given that our calculations will focus on the Tin isotopes, we adopt pairing strengths from 
the G-matrix calculation of \citet{Holt:1998} (see Table II). In particular, it has been shown that shell-model calculations with 
these pairing strengths yield an accurate spectroscopy for the Tin isotopes in the $A\!=\!120\!-\!130$ region. 
\begin{widetext}
\begin{center}
\begin{table}[h]
\begin{tabular}{|l||c|c|c|c|c|}
 \hline
Orbital            &  1g${}_{7/2}$  &  2d${}_{5/2}$  &  2d${}_{3/2}$  &  3s${}_{1/2}$  &  1h${}_{11/2}$  \\  
 \hline
 \hline
1g${}_{7/2}$    &     -0.2463      &     -0.1649      &     -0.1833      &     -0.1460       &       0.2338       \\
2d${}_{5/2}$    &                      &     -0.2354      &     -0.3697      &     -0.1995       &       0.2250       \\
2d${}_{3/2}$    &                      &                      &     -0.2032      &     -0.2485       &       0.1761       \\
3s${}_{1/2}$    &                      &                      &                      &     -0.7244       &       0.1741       \\
1h${}_{11/2}$  &                      &                      &                      &                       &       0.1767       \\
\hline
\end{tabular}
\caption{Pairing strengths $G_{jj'}\!=\!G_{j'\!j}$ (in MeV) for the model space of the Tin isotopes 
($A\!=\!100$ to $A\!=\!132$) as determined from the G-matrix calculation of \citet{Holt:1998}.}
\label{Table2}
\end{table}
\end{center}
\end{widetext}

\section{Results}
\label{Results}

To demonstrate the applicability and utility of the combined RMF+EP approach we now display calculations for the
isotopic chain in Tin: from $A\!=\!100$ to $A\!=\!132$. As mentioned earlier, we regard the doubly magic ${}^{100}$Sn  
nucleus as an inert core and then concentrate on the impact of $nn$-pairing on the $N\!\leq\!32$ neutrons in the valence 
shell. The RMF models listed in Table\,\ref{Table1} are fairly successful in reproducing ground-state properties (such
as binding energies and charge radii) of a variety of nuclei throughout the nuclear chart. However, their predictions for 
some bulk properties of nuclear matter and neutron-star observables differ considerably. For example, the incompressibility
coefficient of symmetric nuclear matter is predicted by FSUGold to be $K_{0}\!=\!230$\,MeV whereas NL3 suggests 
$K_{0}\!=\!271$\,MeV. Moreover, the slope of the symmetry energy $L$---which controls the softening of the GMR
energy along an isotopic chain\,\cite{Piekarewicz:2008nh,Piekarewicz:2009gb,Piekarewicz:2013bea}---is also significantly 
different: $L\!=\!61$\,MeV for FSUGold and $L\!=\!118$\,MeV for NL3. Thus, these two models---one soft and one 
stiff---provide an adequate representative set for the illustration of the method.

\begin{figure}[ht]
\vspace{-0.05in}
\includegraphics[width=0.44\columnwidth,angle=0]{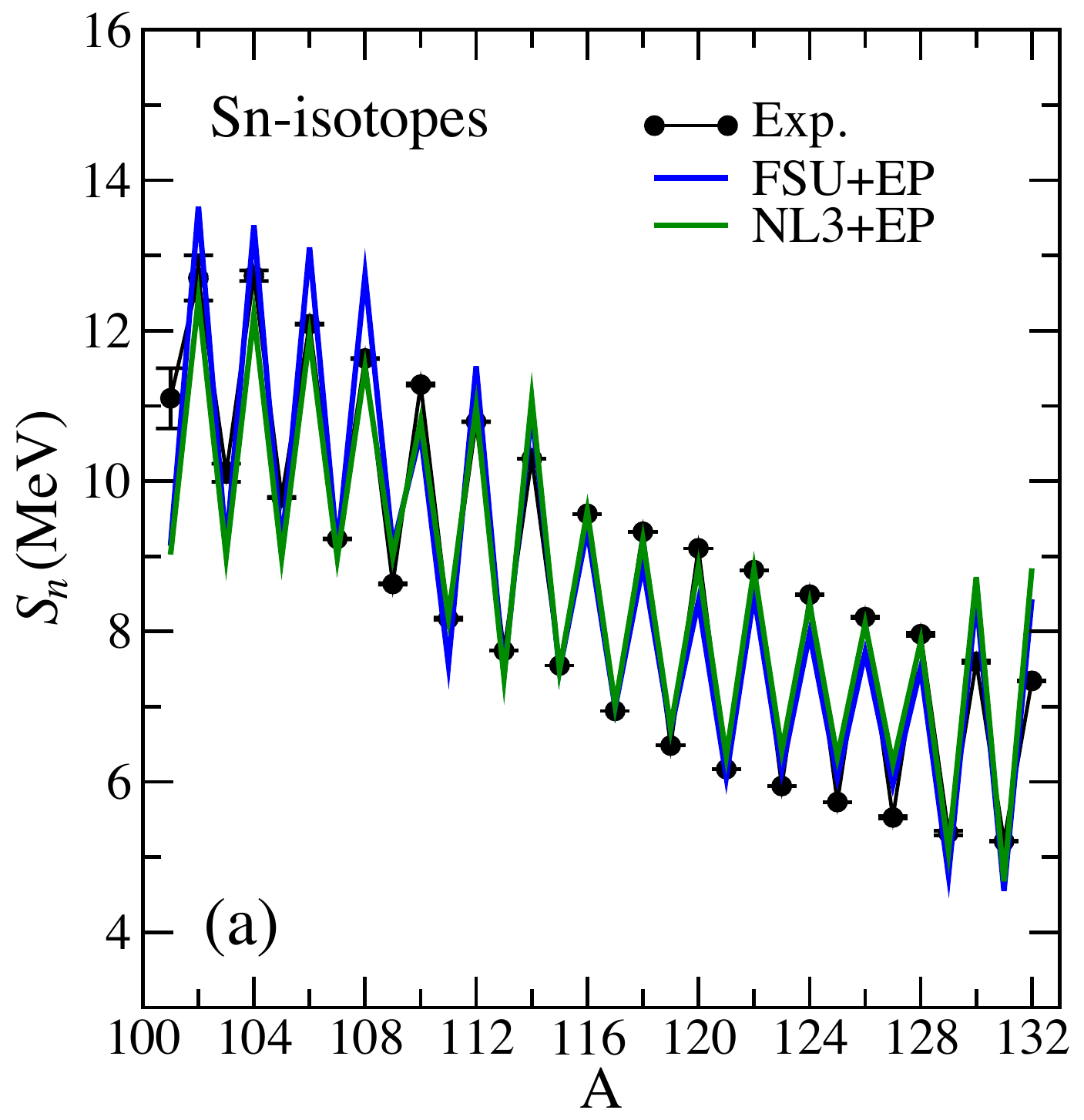}
\hspace{0.5cm}
\includegraphics[width=0.44\columnwidth,angle=0]{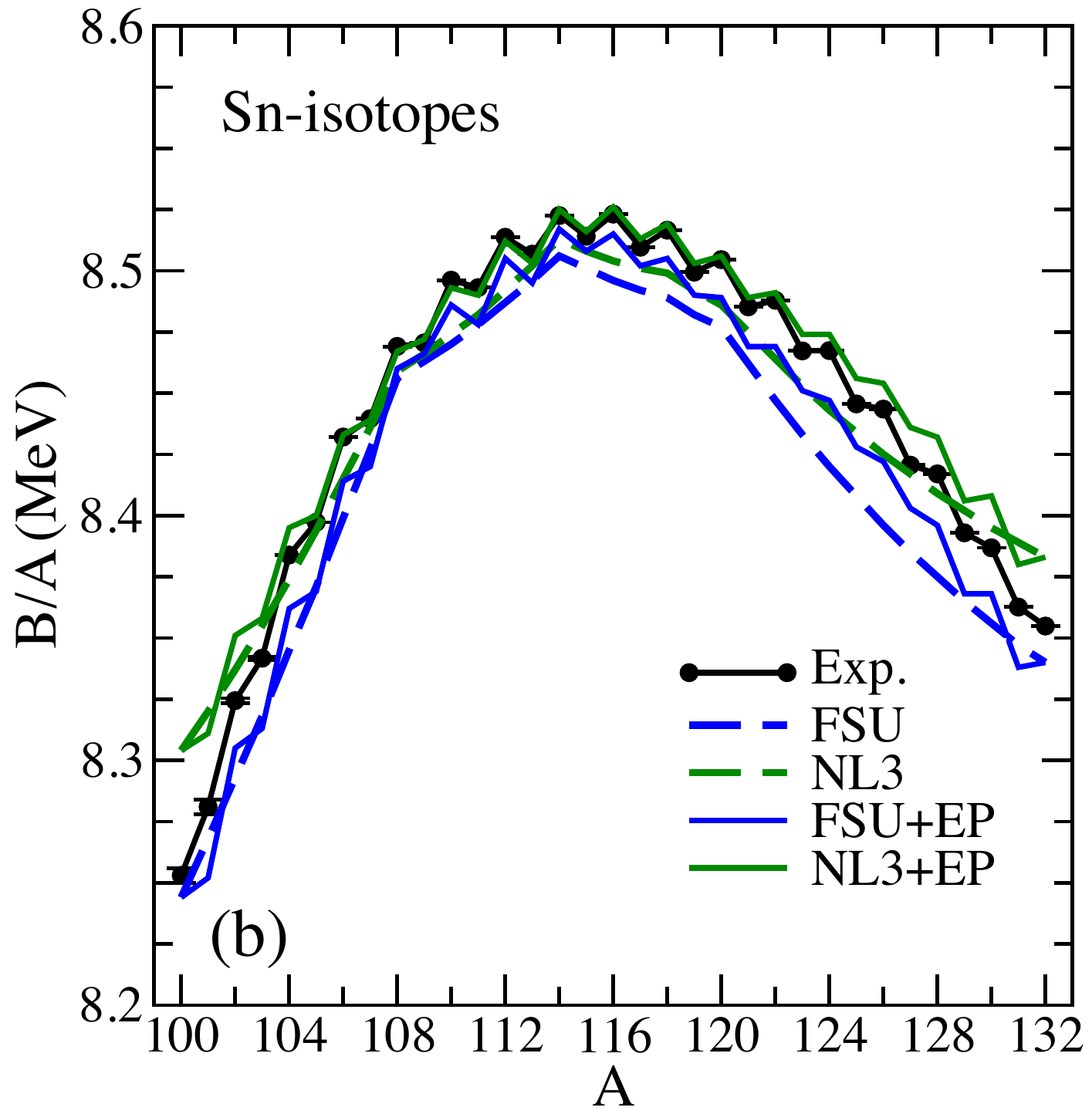}
\caption{(color online) Odd-even staggering of the one-neutron separation energy (left panel) and binding energy 
per nucleon (right panel) along the isotopic chain in Tin as predicted by the FSUGold\,\cite{Todd-Rutel:2005fa} 
and NL3\,\cite{Lalazissis:1996rd} models---supplemented with an exact treatment of pairing corelations. 
Experimental results are from Ref.\,\cite{NNDC}.} 
\label{Fig1}
\end{figure}

On the left-hand panel of Fig.\,\ref{Fig1} we show one of the classical signatures of pairing correlations: the odd-even 
staggering of the neutron separation energy along the isotopic chain in Tin; experimental data are from the compilation 
given in Ref.\,\cite{NNDC}. Note that for odd-A nuclei the unpaired neutron is placed in the single-particle orbital that 
reproduces the experimental angular momentum and parity of the ground state\,\cite{NNDC}. In the case of even-A 
nuclei all neutrons are paired. This odd-even difference gives rise to the characteristic staggering observed in the 
one-neutron separation energy. That is, in the case of an even-A nucleus, one must provide the energy necessary to 
break the pair before the neutron can be excited into the continuum. In contrast, for odd-A nuclei there is no additional 
cost associated with breaking a pair. Fig.\,\ref{Fig1} suggests that the energy required to break a pair is about 3\,MeV, 
which is the typical strength associated with the residual interaction. Moreover, it can be seen that for the stable 
A=112-124 nuclei the results are in good agreement with experiment. However, deviations of about 1MeV seem to 
emerge at the two ends of the isotopic chain. We would like to point out that the ground-state spin of some of those 
unstable odd-A nuclei is uncertain. In addition, if the neutron 1g${}_{9/2}$ orbital (which so far has been assumed inert) 
is incorporated into the pairing calculation, the discrepancies in the light isotopes are expected to disappear. Similarly, 
we expect that the predictions for the heavier isotopes will improve as one includes higher neutron orbitals. On the 
right-hand panel of Fig.\,\ref{Fig1} we display the binding energy per nucleon along the isotopic chain. Many of the same 
features already evident in the one-neutron separation energy are also manifest in this observable. However, in this case 
we also display the predictions from the RMF models \emph{without} pairing correlations. As expected---and in sharp 
contrast to the experimental data---the predicted A-dependence is smooth and devoid of the ``zigzag'' structure.

\begin{figure}[ht]
\vspace{-0.05in}
\includegraphics[width=0.5\columnwidth,angle=0]{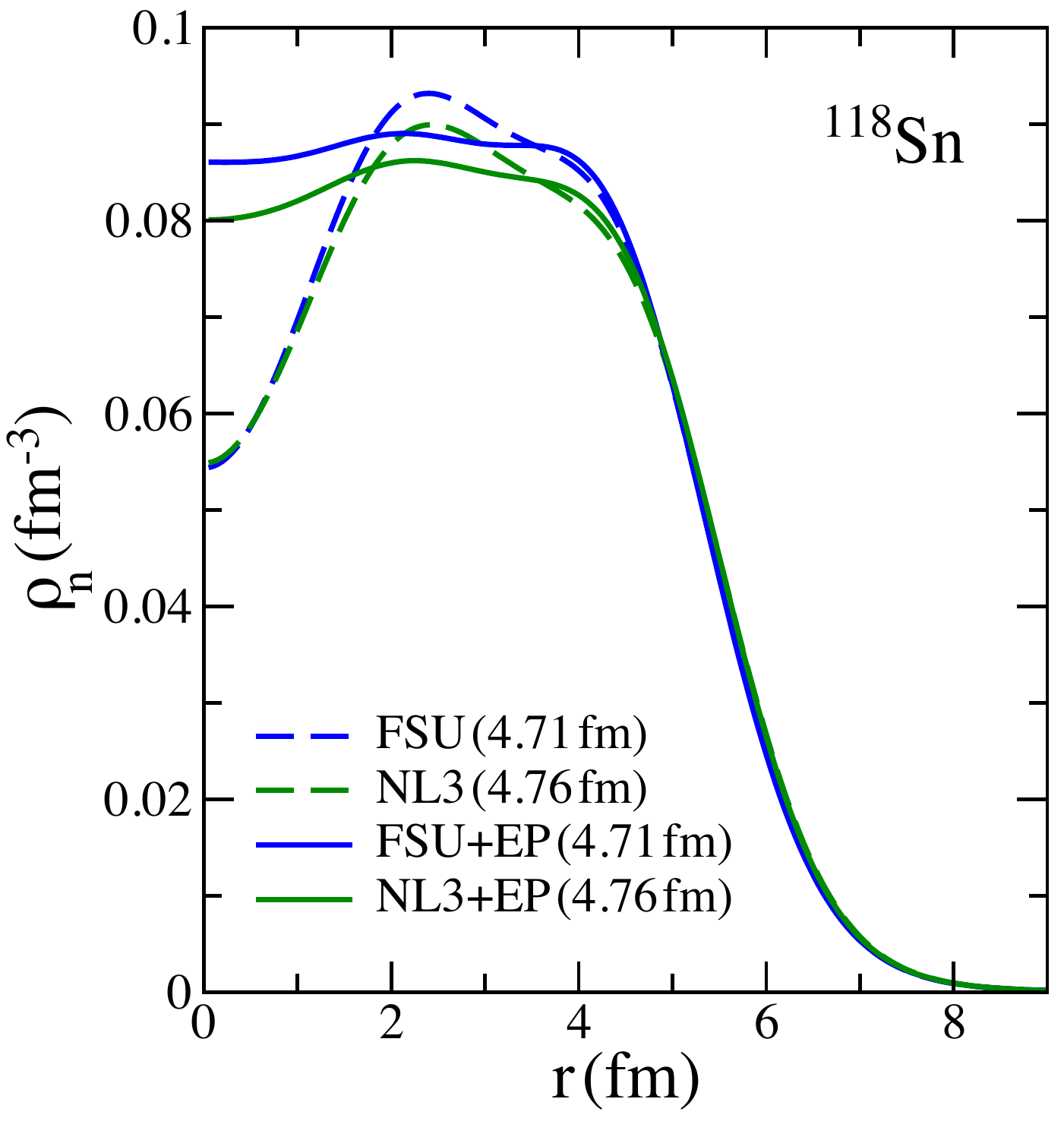}
\caption{(color online) Neutron density of ${}^{118}$Sn as predicted by the  FSUGold\,\cite{Todd-Rutel:2005fa} and 
NL3~\cite{Lalazissis:1996rd} models. Predictions are displayed with (solid lines) and without (dashed lines) pairing
correlations. Quantities enclosed in parenthesis represent the model predictions for the root-mean-square radius.} 
\label{Fig2}
\end{figure}

Another critical signature of pairing correlations is the fractional occupancy of the single-particle orbits. Thus, the
predicted fractional occupancies $\langle N_{j}\rangle$ for the five neutron orbitals forming the valence space are 
displayed in Table\,\ref{Table3} for all stable, even-A Tin isotopes. Predictions are presented for FSUGold and 
(separated by a ``/") for NL3. As shown in the table, pairing correlations can modify the occupancies by as much 
as one neutron relative to the naive mean-field expectations. In particular, such changes may have a significant 
impact on the novel ``bubble" structure and concomitant quenching of the spin-orbit splitting of low-$j$ 
orbitals\,\cite{Todd-Rutel:2004tu,Gaudefroy:2006zz,Piekarewicz:2006by,Grasso:2009zza}. Indeed, in Fig.\,\ref{Fig2} 
we exhibit the ground-state neutron density of ${}^{118}$Sn as predicted by both RMF models with and without the 
inclusion of pairing correlations. In the extreme mean-field limit, ${}^{118}$Sn consists of filled 1g${}_{7/2}$, 2d${}_{5/2}$, 
and 2d${}_{3/2}$ orbitals. In particular, the absence of 3s${}_{1/2}$ neutrons yields the bubble structure (manifested 
as a central depression) in the neutron density. In turn, such a central depression leads to a modification 
of the spin-orbit potential that results in a quenching of the spin-orbit splitting between the 2p${}_{3/2}$ and 2p${}_{1/2}$ 
proton orbitals: 0.85\,MeV for FSUGold and 0.77\,MeV for NL3. However, the 3s${}_{1/2}$ orbital lies within 
$\sim\!0.5$\,MeV of the 2d${}_{3/2}$ orbital, so the mixing induced by the pairing interaction is very efficient
(see Table\,\ref{Table3}). Indeed, with more than one neutron transferred to the 3s${}_{1/2}$ orbital, the bubble
structure of ${}^{118}$Sn disappears entirely. Moreover, the 2p${}_{3/2}$-2p${}_{1/2}$ spin-orbit splitting increases 
by more than 50\% to 1.33\,MeV for both FSUGold and NL3. Note, however, that the
occupancy of the 3s${}_{1/2}$ orbital in both ${}^{112}$Sn and ${}^{114}$Sn remains small so their bubble structure
is preserved---although not as pronounced as in the case of ${}^{118}$Sn.

\begin{widetext}
\begin{center}
\begin{table}[h]
\begin{tabular}{|l||c|c|c|c|c|c|c|}
\hline
                       &  ${}^{112}$Sn  &  ${}^{114}$Sn  &  ${}^{116}$Sn  &  ${}^{118}$Sn  &  ${}^{120}$Sn  &  ${}^{122}$Sn  &  ${}^{124}$Sn  \\
\hline
\hline
1g${}_{7/2}$    &    7.80/7.58    &    7.89/7.80    &    7.88/7.80    &    7.89/7.81    &    7.92/7.84    &    7.91/7.85    &    7.92/7.87    \\  
2d${}_{5/2}$    &    3.64/3.72    &    5.46/5.42    &    5.54/5.49    &    5.66/5.58    &    5.79/5.68    &    5.79/5.72    &    5.81/5.77    \\
2d${}_{3/2}$    &    0.28/0.32    &    0.33/0.36    &    1.24/1.30    &    2.31/2.24    &    3.42/3.01    &    3.52/3.30    &    3.62/3.50    \\
3s${}_{1/2}$    &    0.08/0.09    &    0.11/0.12    &    0.86/0.72    &    1.38/1.15    &    1.77/1.52    &    1.81/1.66    &    1.84/1.76    \\
1h${}_{11/2}$  &    0.20/0.29    &    0.21/0.30    &    0.47/0.70    &    0.76/1.21    &    1.11/1.96    &    2.97/3.46    &    4.80/5.11    \\
\hline
\end{tabular}
\caption{Single-particle occupancies $\langle N_{j}\rangle$ of the orbitals in the valence shell for the stable even-even Sn isotopes. 
Results are presented for FSUGold\,\cite{Todd-Rutel:2005fa} and (separated by a ``/'') for NL3\,\cite{Lalazissis:1996rd}.}
\label{Table3}
\end{table}
\end{center}
\end{widetext}

We finish this section by addressing the role of pairing correlations in explaining the softness of the Tin 
isotopes\,\cite{Li:2007bp,Li:2010kfa}. The study of pairing correlation on the GMR energies of the Tin 
isotopes dates back to the work of  \citet{Civitarese:1991} in the early 1990s. By employing a 
quasiparticle-RPA formalism, the authors reported a small shift of about 100 to 150 keV in the monopole 
energies due to pairing correlations. Recently, the role of pairing correlations has been revisited as a 
possible mechanism to soften the GMR energies of these superfluid 
nuclei\,\cite{Li:2008hx,Khan:2009xq,Khan:2009ih,Khan:2010mv,Vesely:2012dw}. Thus, it seems
natural to examine this critical issue within the context of the RMF+EP approach introduced here.

To investigate the effect of pairing correlations on the monopole energies we invoke the newly developed 
constrained-RMF approach introduced in Ref.\,\cite{Chen:2013tca}. As already alluded in Sec.\,\ref{CRMFT}, 
the convenience of the constrained approach stems from the accurate and efficient estimation of GMR 
energies without the need to generate the full distribution of monopole strength. Indeed, as indicated in 
Eqs.\,(\ref{EWSR}) and\,(\ref{IEWSR}), GMR energies may be computed directly from the mean-square radius 
of the \emph{ground-state} distribution. Recently, excellent results were obtained as the CRMF approach was 
compared against the predictions from a relativistic RPA calculation\,\cite{Chen:2013tca}.
To examine the impact of pairing correlations on the GMR energies, the ground-state densities that serve 
as the input for the constrained approach will now be calculated using the RMF+EP formalism. A first 
glance at Fig.\,\ref{Fig2} may suggest that pairing correlations could have a dramatic effect on the GMR 
energies. However, upon closer inspection one realizes that it is the mean-square radius of the ground-state 
density that is of relevance to the GMR energies. Hence, the $r^{4}$ weighting of the density, namely, 
\begin{equation}    
  \langle r^2\rangle = \frac{4\pi}{A} \int r^{4}\rho(r)\,dr \;,
 \label{RSquare}
\end{equation}
washes out the dramatic effect observed in the central density. Indeed, we find no modification to the mean-square 
radius from pairing correlations; the root-mean-square radius of the neutron density of ${}^{118}$Sn is displayed 
in Fig.\,\ref{Fig2}.

\begin{figure}[ht]
\vspace{-0.05in}
\includegraphics[width=0.5\columnwidth,angle=0]{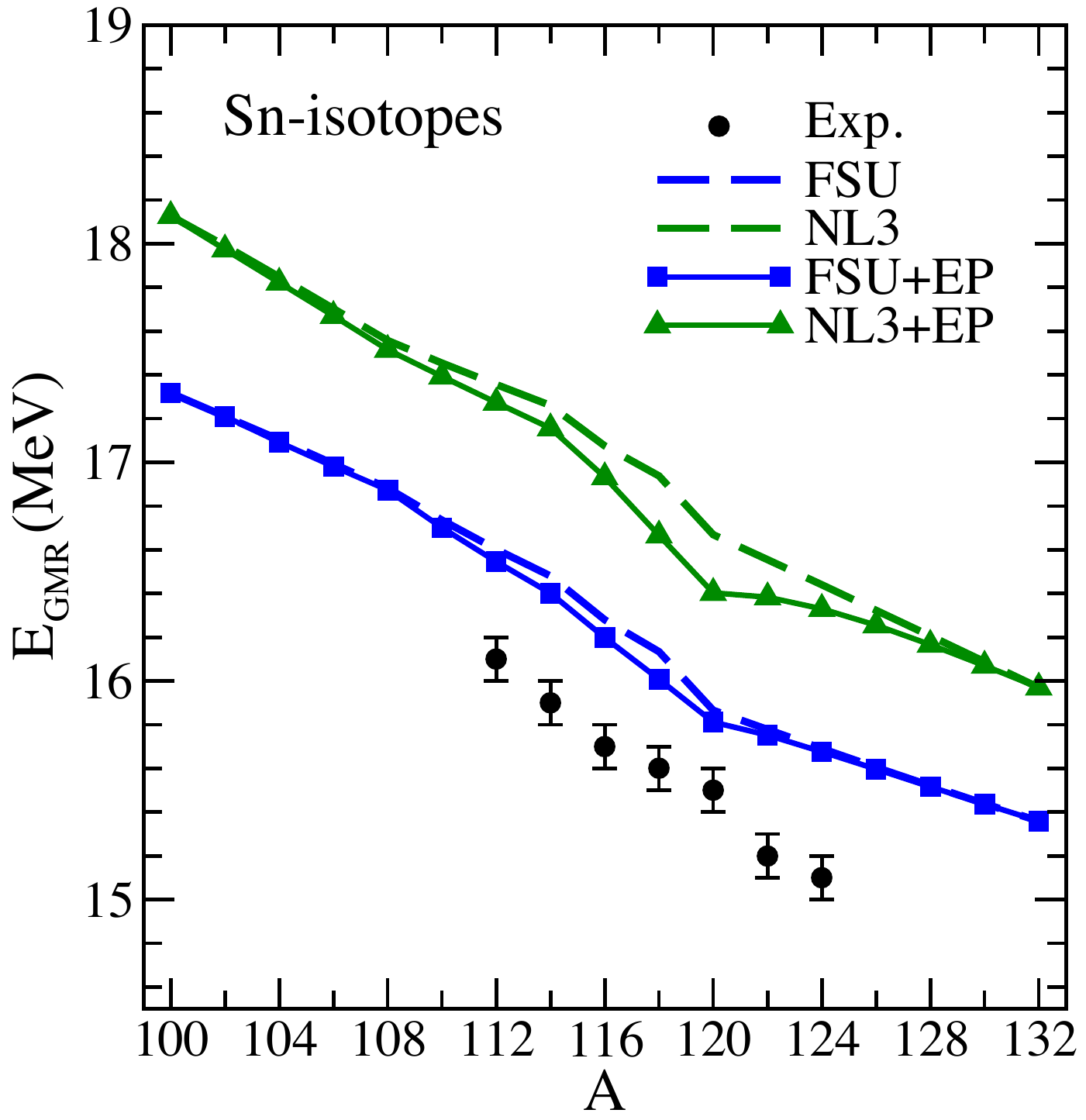}
\caption{(color online) Giant monopole energies in the even-even Sn isotopes with and without pairing
correlations compared against the experimental results of \citet{Li:2007bp}. Predictions from both 
FSUGold\,\cite{Todd-Rutel:2005fa} and NL3\,\cite{Lalazissis:1996rd}---with and without 
pairing correlations---overestimate the experimental data.}
\label{Fig3}
\end{figure}

GMR energies for the whole isotopic chain in Tin---from A=100 to A=132---are displayed in Fig.\,\ref{Fig3}
alongside the experimental results for the stable even-A isotopes\,\cite{Li:2007bp}. As the number of 
neutrons increases, the GMR energy decreases monotonically. In this regard two points are worth 
emphasizing: (a) the value of the centroid energy in ${}^{112}$Sn and (b) the softening of the mode as a 
function of $A$. First, given that the neutron excess in ${}^{112}$Sn is small, the value of its centroid energy 
is mostly sensitive to the incompressibility coefficient of symmetric nuclear matter. This is clearly reflected 
in the model predictions; recall that $K_{0}\!=\!230$\,MeV for FSUGold and $K_{0}\!=\!271$\,MeV for NL3.
However, even the significantly softer FSUGold model overestimates the centroid energy in ${}^{112}$Sn 
by about 0.3 MeV. Second, the experiment suggests a very rapid softening that is not reproduced by 
either of the models. Note that the falloff with $A$ is largely controlled by the slope of the symmetry
energy $L$\,\cite{Piekarewicz:2013bea}. Thus, whereas the falloff predicted by NL3 may indeed be
slightly faster than that of FSUGold, it is clearly nowhere as fast as required by the experiment. 
Thus, although we have gone beyond a mean-field-plus-RPA description, Fig.\,\ref{Fig3} indicates that 
the impact of pairing correlations on the GMR energies is fairly small---especially in the case of FSUGold. 
Indeed, the largest correction due to pairing is about 275\,keV for NL3 and about 125\,keV for FSUGold. Given 
that in the constrained approach the GMR energy is driven by the mean-square radius of the nuclear distribution,
it hardly comes as a surprise that pairing correlations play a minor role. Moreover, although both RMF
models have been accurately calibrated, NL3 predicts a valence spectrum that is in general more 
compressed than the one predicted by FSUGold. Thus, pairing correlations are more quenched 
in FSUGold than in NL3. In addition, from ${}^{102}$Sn to ${}^{114}$Sn the valence neutrons reside in the 
1g${}_{7/2}$ and 2d${}_{5/2}$ orbitals---which are relatively well separated from the 2d${}_{3/2}$, 3s${}_{1/2}$, 
and 1h${}_{11/2}$ orbitals. Thus, pairing correlations play a minor role in populating the three higher orbitals 
(see Table\,\ref{Table3}). However, once the higher orbitals start to be populated, pairing correlations become
very efficient at redistributing $nn$-pairs, especially among the quasi-degenerate 2d${}_{3/2}$ and 
3s${}_{1/2}$ orbitals. This is particularly true in the middle of the shell, namely, from ${}^{116}$Sn to ${}^{120}$Sn. 
After that the effect from pairing correlations weakens because transitions to the partially occupied 1h${}_{11/2}$ 
orbital become Pauli blocked. Note that contrary to the predictions of Ref.\,\cite{Khan:2009xq}---where constrained 
HFB calculations using Skyrme functionals and a zero-range surface pairing force were performed---we do not 
observe a rapid stiffening of the mode (in the form of a prominent peak) as one reaches the doubly-magic nucleus 
${}^{132}$Sn.

\section{Conclusions}
\label{Conclusions}

In this work we introduced a novel hybrid approach to compute the properties of open-shell nuclei. The method consists
of a relativistic mean-field approximation supplemented with an exact treatment of pairing correlations. One of the 
major advantage of the exact treatment is that it conserves the number of particles. This avoids any reliance on 
complicated prescriptions that must be used to project out the correct number of particles. Moreover, the EP approach 
works well even for nuclei with typical single-particle-energy spacings greater than the pairing strengths---a limit in which 
both BCS and HFB tend to fail. Finally, the combined RMF+EP approach is simple to implement. One 
starts by computing the single-particle spectrum using a traditional mean-field approach. Once the valence shell is identified, 
then one passes the relevant single-particle energies to the exact-pairing routine---which redistributes pairs among 
the valence orbitals. The newly-obtained fractional occupancies then generate a new set of baryon densities that 
ultimately yield an updated single-particle spectrum. This updated spectrum now serves as input to the exact-pairing routine 
and the procedure is repeated until self-consistency is achieved. 

The utility and applicability of the RMF+EP approach was demonstrated using the long chain of Tin isotopes as an
example. Predictions for both ground-state properties and GMR energies were compared against experimental
results. In the case of ground-state properties, results were presented for the characteristic odd-even staggering of 
the one-neutron separation energy and binding energies across the full isotopic chain: from ${}^{100}$Sn 
to ${}^{132}$Sn. We find that our predictions compare very favorably against the experimental results. In addition,
we presented results for the single-particle occupancies of the relevant neutron orbitals and investigated their impact
on the allegedly nuclear bubble structure. In particular, we concluded that the bubble structure observed in the 
``mean-field'' neutron density of ${}^{118}$Sn is completely eliminated with the inclusion of pairing correlations. 
However, we find that the bubble structure of both ${}^{112}$Sn and ${}^{114}$Sn, although not as pronounced as 
in the case of ${}^{118}$Sn, is still maintained. 

We also investigated the role of pairing correlations on the possible softening of the GMR energies in the Tin isotopes. 
To do so, we adopted the recently developed constrained-RMF framework that has been shown to accurately reproduce 
GMR energies obtained with the more sophisticated RPA approach. The great merit of the constrained approach is that 
GMR energies can be accurately and efficiently computed from the mean-square radius of the \emph{ground-state} density 
distribution. Thus, in this work we investigated the role of pairing correlations by supplying the CRMF approach with 
densities computed in the RMF+EP framework. As has been extensively documented, models that reproduce the GMR 
energies in both ${}^{90}$Zr and ${}^{208}$Pb overestimate the corresponding monopole energies along the isotopic chain 
in Tin. Given that pairing correlations have been proposed as a possible solution to the softening of these superfluid nuclei,
GMR energies along the isotopic chain in Tin---from ${}^{100}$Sn to ${}^{132}$Sn---were computed using the combined 
RMF+EP formalism. We concluded, as many have done before us, that pairing correlations provide (if at all!) a very mild 
softening of the mode. Within the constrained approach the explanation for this behavior is rather simple. Whereas pairing 
correlations modify the ground-state density distribution, most of these modifications are limited to the nuclear interior. 
Given that the constrained energy is driven by the mean-square radius of the density distribution---which is largely 
insensitive to the nuclear interior---pairing correlations play a minor role in the softening of the mode. Thus, we conclude 
that pairing correlations can not be the explanation behind the question of ``\emph{why is Tin so soft?''}

In summary, we have introduced a novel RMF+EP approach to compute ground-state properties and collective 
excitations of open-shell nuclei. The approach is elegant and straightforward, and its implementation fast and 
reliable. Moreover, particle-number conservation is strictly maintained, so the approach is not hindered by
complicated projection prescriptions required in other formulations. We are confident that the combined 
RMF+EP approach introduced here provides a simple and powerful framework for the exploration of the limits 
of nuclear existence, such as in the study of superheavy nuclei and of nuclei near the drip lines.

\begin{acknowledgments}
This work was supported in part by the United States Department of Energy under grants 
DE-FG05-92ER40750 and DE-SC0009883.
\end{acknowledgments}

\appendix*
\section{A Toy Model of the Exact-Pairing Algorithm}

To illustrate how the exact pairing algorithm is implemented we present here a toy model consisting
of 4 neutrons residing in 2 single-particle orbitals, such as $2d_{3/2}$ (label 1) and $3s_{1/2}$ (label 2). 
This example may reflect the simplified situation in which ${}^{114}$Sn may be assumed as an 
inert core and one is interested in studying the structure of ${}^{118}$Sn---particularly the occupancy of
the quasi-degenerate $2d_{3/2}$ and $3s_{1/2}$ orbitals. Assuming that all 4 neutrons are paired, 
{\sl i.e.,} both orbitals have seniority zero, there are only 2 allowed configurations:
\begin{equation}
 |a\rangle\!=\!|N_{1}\!=\!4,N_{2}\!=\!0\rangle \quad{\rm and}\quad 
 |b\rangle\!=\!|N_{1}\!=\!2,N_{2}\!=\!2\rangle\,.
\end{equation}
Matrix elements of the pairing Hamiltonian in the seniority basis are now obtained from the general
expressions derived in Ref.\,\cite{Volya:2001}. For example, the diagonal matrix elements of the
pairing Hamiltonian are given by
\begin{equation}
\langle N_{1},N_{2} | H | N_{1},N_{2} \rangle = 
 \sum_{j=1}^{2} \left[\epsilon_j N_j + \frac{G_{jj}}{4} N_j (2\Omega_j\!-\!N_j\!+\!2) \right],
\end{equation}	 
where $\epsilon_{j}$ are single-particle energies, $G_{jj}$ are (diagonal) pairing strengths, and $\Omega_j$ 
is the pair degeneracy of orbital $j$; $\Omega_{1}\!=\!2$ and $\Omega_{2}\!=\!1$. Similarly, the off-diagonal 
matrix elements are given by
\begin{equation}
 \langle N_{1}\!+\!2,N_{2}\!-\!2 | H | N_{1},N_{2}  \rangle = 
 \frac{G_{12}}{4} \sqrt{N_{2} (2\Omega_{2}\!-\!N_{2}\!+\!2) (2\Omega_1\!-\!N_1) (N_1\!+\!2) },
\end{equation}
where now $G_{12}$ is the pair-transfer strength. By assuming a constant pairing strength 
$G_{jj'}\!\equiv\!-g$ (with $g\!>\!0$) the $2\!\times\!2$ pairing Hamiltonian takes the following 
simple form:
\begin{equation}
 H  = \left( \begin{array}{cc} 
         4\epsilon_{1}-2g & -\sqrt{2}g \\
         -\sqrt{2}g & 2\epsilon_{1}+2\epsilon_{2}-3g 
         \end{array}\right) =
         E\mathds{1}+\left( \begin{array}{cc} 
         -\epsilon & -\sqrt{2}g \\
         -\sqrt{2}g & \epsilon
         \end{array}\right)\,,
\end{equation} 
where we have introduced the following definitions (with $\Delta\!\equiv\!\epsilon_{2}\!-\!\epsilon_{1}$):
\begin{equation}
     E \equiv 4\epsilon_{1}+\Delta-\frac{5}{2}g \quad{\rm and}\quad
     \epsilon \equiv \Delta - \frac{1}{2}g \,.     
\end{equation}
Diagonalizing the pairing Hamiltonian yields the following value for the ground-state energy and for 
its corresponding eigenstate: 
\begin{subequations}
 \begin{align}
 & E_{0}=E-\xi\equiv E-\sqrt{\epsilon^{2}+2g^{2}}\,, \\
 & |E_{0}\rangle = \sqrt{\frac{\xi+\epsilon}{2\xi}}\,\Big|N_{1}\!=\!4,N_{2}\!=\!0\Big\rangle
                        + \sqrt{\frac{\xi-\epsilon}{2\xi}}\,\Big|N_{1}\!=\!2,N_{2}\!=\!2\Big\rangle \;.
  \end{align}                       
\end{subequations}
In a mean-field calculation without pairing correlations, the 4 neutrons would occupy the lowest 2$d_{3/2}$ orbital
with the lowest energy being equal to $E_{0}\!=\!4\epsilon_{1}$. Pairing correlations reduce the ground-state energy 
at the expense of redistributing the 4 neutrons among the 2 valence orbitals. In this way the average occupancy of 
the 2 orbitals becomes
\begin{equation}
     \langle N_{1} \rangle = 3 +\frac{\epsilon}{\xi}
     \quad{\rm and}\quad
     \langle N_{2} \rangle = 1 - \frac{\epsilon}{\xi} \,.
\end{equation}
Note that the fractional occupancies of the single-particle orbits depend exclusively on the ratio of $\Delta/g$.
In Table\,\ref{Table4} we list (properly scaled) ground-state energies as well as fractional occupancies for the
two valence orbitals. At values of $\Delta/g\!\simeq\!1$ the ground-state is well correlated and the occupancy
of the lowest orbital gets significantly depleted. In contrast, for $\Delta/g\!\gg\!1$, the single-particle gap is
significantly larger than the pairing strength and the occupancy of the lowest state returns to its mean-field
value of 4.
\begin{widetext}
\begin{center}
\begin{table}[h]
\begin{tabular}{|c|c|c|c|}
 \hline
$\Delta/g$   &   $(E_{0}\!-\!4\epsilon_{1})/g$ & $\langle N_{1} \rangle$ & $\langle N_{2} \rangle$ \\
 \hline
 \hline
 1/2  & -3.414 & 3.000 & 1.000 \\
 1     & -3.000 & 3.333 & 0.667 \\
 2     & -2.562 & 3.728 & 0.272 \\
 4     & -2.275 & 3.927 & 0.073 \\ 
 8     & -2.132 & 3.983 & 0.017 \\
16    & -2.064 & 3.996 & 0.004 \\
 \hline
\end{tabular}
\caption{Ground-state energy of the pairing Hamiltonian and corresponding single-particle occupancies 
for different values of $\Delta/g$.}
\label{Table4}
\end{table}
\end{center}
\end{widetext}

\bibliography{./RMFEP.bbl}

\end{document}